\documentclass{pasa}%
\newcommand{\kms}{km\,s$^{-1}$}
\newcommand{\siii}{[S\,{\sc iii}]}
\newcommand{\arc}{$^{\prime\prime}$}

\pdfoutput=1

\usepackage{graphicx}

\title[Outflows in NGC\,5643]{Outflows in the Seyfert 2 galaxy NGC\,5643
 traced by the [SIII] emission}

\author[Riffel, Hekatelyne \& Freitas]{Rogemar A. Riffel$^1$, C. Hekatelyne$^1$, Izabel C. Freitas$^{1,2}$ 
\affil{$^1$Universidade Federal de Santa Maria, Departamento de F\'\i sica, CCNE,  97105-900, Santa Maria, RS, Brazil\\
$^{2}$Universidade Federal de Santa Maria, Col\'egio Polit\'ecnico, 97105-900, Santa Maria, RS, Brazil }
}%

\jid{PASA}
\doi{10.1017/pas.\the\year.xxx}
\jyear{\the\year}

\usepackage{aas_macros}
\usepackage{hyperref} 
\hypersetup{colorlinks,citecolor=blue,linkcolor=blue,urlcolor=blue}

\hypersetup{draft}

\begin{document}

\begin{frontmatter}
\maketitle

\begin{abstract}

We use Gemini Multi-Object  Spectrograph (GMOS) Integral Field Unit (IFU) observations of the inner 285$\times$400\,pc$^2$ region of the Seyfert 2 galaxy NGC\,5643 to map the \siii$\lambda9069$ emission-line flux distribution and kinematics, as well as the stellar kinematics, derived by fitting the Ca\,{\sc ii}$\lambda\lambda\lambda$8498,8542,8662 triplet,  at a spatial resolution of 45\,pc.  The stellar  velocity field shows regular rotation, with a projected velocity of 100~\kms\ and kinematic major axis along Position Angle $PA=-36^\circ$. A ring of low stellar velocity dispersion values ($\sim$70\,\kms), attributed to young/intermediate age stellar populations,  is seen surrounding the nucleus with radius of 50~pc. We found that the \siii\ flux distribution shows an elongated structure along the east-west direction and its kinematics is dominated by outflows within a bi-cone at an ionized gas outflow rate of 0.3\,M$_\odot$\,yr$^{-1}$. In addition, velocity slices across the \siii$\lambda9069$ emission-line reveal a kinematic component attributed to rotation of gas in the plane of the galaxy.

\end{abstract}

\end{frontmatter}
\section{INTRODUCTION }
\label{sec:intro}

\begin{figure*}[t]
\begin{center}
\includegraphics[width=0.8\textwidth]{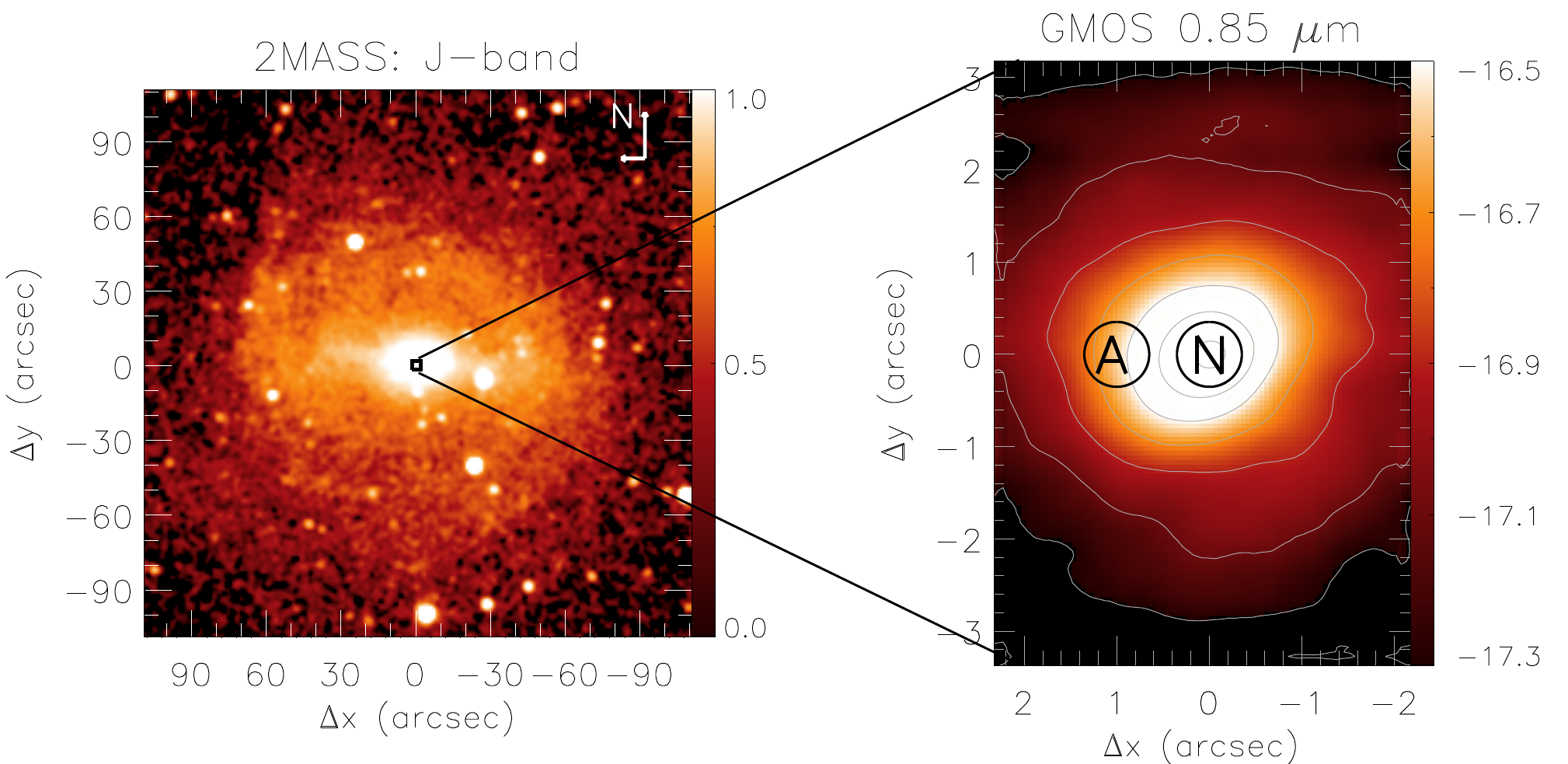}
\includegraphics[width=0.8\textwidth]{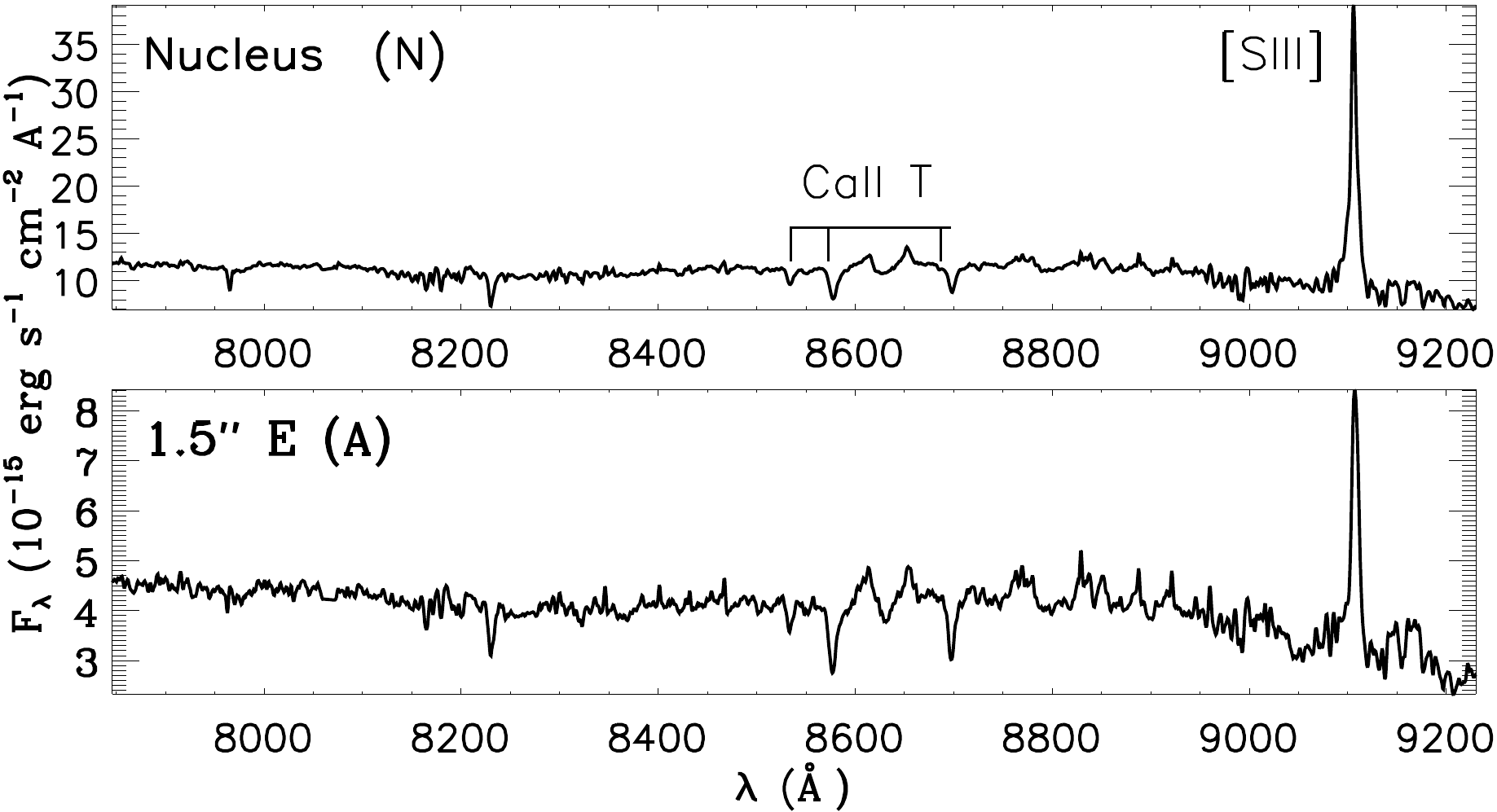}
\caption{Top panels -- Left: Large scale J-band image of NGC\,5643 from 2MASS \citep{2mass}. The color bar shows the flux in arbitrary units. Right: 0.85\,$\mu$m continuum image obtained from the GMOS datacube, by averaging the fluxes within a 300\,\AA\ spectral range. The color bar shows the fluxes in logarithmic units of erg\,s$^{-1}$\,cm$^{-2}$\,\AA$^{-1}$\,spaxel$^{-1}$. Bottom panels show the spectra extracted within the circular apertures of 0.7\arc\ diameter centred at the nucleus (N) and at 1.5\arc\ of it (A). \label{large}}
\end{center}
\end{figure*}

The unified model of Active Galactic Nuclei (AGN) postulates the existence of ionization bi-cones, delineating the Narrow Line Region \citep[NLR;][]{antonucci93,up95}. Within the bi-cone, winds from the accretion disk are expected to be observed.  Indeed these conical shaped structures were seen in early ground based images \citep{wilson94}. However, later, [O\,{\sc iii}]$\lambda5007$  high-resolution narrow-band images obtained with the Hubble Space Telescope (HST) showed that the conical morphology is not as common as previously thought \citep[e.g.][]{schmitt03}. \citet{fischer13} used HST long-slit observations of a sample of 48 AGN to map the [O\,{\sc iii}] kinematics and found that only 35\,\% of the galaxies with extended NLR show outflows. Integral Field Spectroscopy (IFS) on 8-10 m telescopes is a powerful tool to map the central region of active galaxies, as they provide the spatial and spectral resolutions and spatial coverage, needed to properly map the NLR kinematics and geometry \citep[e.g.][]{eso428,barbosa09,mrk79,n5929,medling15,lena16,fischer17,thomas17,dominika17,freitas18}.

We present near-infrared (near-IR) IFS of the inner  $5^{\prime\prime}\times7^{\prime\prime}$ region of the Seyfert 2 galaxy NGC\,5643, obtained with the Gemini Multi-Object Spectrograph (GMOS).  NGC\,5643 is a widely studied spiral galaxy, morphologically classified as SBc and located at a distance of 11.8\,Mpc, for which 1\arc\ corresponds to 57\,pc at the galaxy \citep{dv76,fischer13}. It harbors a Seyfert 2 nucleus and its optical spectra shows high ionization lines \citep{sandage78,philips83}. 

Ground-based [O\,{\sc iii}]$\lambda5007$ and H$\alpha$+[N\,{\sc ii}]$\lambda\lambda6548,84$ narrow-band images of NGC\,5643 show an elongated structure with size of 30$^{\prime\prime}$ (1.9\,kpc) aligned along the east-west direction, with higher excitation gas observed to the east of the nucleus \citep{schmitt94}. High resolution narrow-band images obtained with the HST have confirmed the presence of higher excitation gas to the east of the nucleus and reveal a well defined one-sided cone with vertex at the location of the continuum emission peak and oriented to the east \citep{simpson97}. Such structure is also detected in soft X-ray images \citep{bianchi06}. 
\citet{fischer13} used HST narrow-band images and long-slit spectroscopy of the NLR of NGC\,5643  and found that the [O\,{\sc iii}] kinematics is consistent with outflows within a cone oriented along Position Angle $PA=80^\circ$, with inclination of 25$^\circ$ and maximum opening angle of 55$^\circ$. 

\citet{cresci15} used high quality IFS of NGC\,5643, obtained with the MUSE instrument on the Very Large Telescope, to map the gas ionization and kinematics in the inner 25$^{\prime\prime}\times$10$^{\prime\prime}$  region. They found a double-sided ionization cone along the east-west direction and argue that the ionized gas kinematics at the center of the ionization cone is consistent with outflows, based on the detection of a blueshifted assymetric wing of the [O\,{\sc iii}]$\lambda5007$ emission line, with projected velocity of up to $-450$\,\kms.  In addition, they found that the outflow points in the direction of two star forming regions, and suggest that these regions are due to a positive feedback induced by the gas compression by the outflowing gas.

Near-IR IFS of the inner $8^{\prime\prime}\times8^{\prime\prime}$ region of NGC\,5643,  obtained with the Very Large Telescope (VLT), reveals that the Br$\gamma$ emission-line shows a similar flux distribution to the optical lines, while the H$_2 1-0\,S(1)$ flux distribution presents two clear spiral arms one to the northwest and another to the southeast \citep{davies14,menezes15}. The H$_2$ velocity field shows kinematic structures associated the spiral arms seen in the H$_2$ flux distribution, and consistent  with gas flows towards the nucleus along a bar. To the northeast, the H$_2$ kinematics is consistent with outflows of molecular gas excited around the edge of the ionization cone.  These authors found also that the Br$\gamma$ emission traces the ionization cone with velocities of up to 150\,\kms. The stellar velocity field of the central region of NGC\,5643, derived by fitting the K-band CO absorption band heads, shows regular rotation with the line of nodes oriented along $PA=-39^\circ$ and projected velocity amplitude of $\sim$100\,\kms\ \citep{hicks13,davies14}.

Recently, \citet{ah18} presented high resolution $^{12}$CO(2--1) line and  232\,GHz continuum observations of NGC\,5643 obtained with the Atacama Large Millimeter/submillimeter Array (ALMA) with spatial resolutions of 9--21\,pc. The CO intensity map shows a two-arm nuclear spiral extending to up to $\sim$10$^{\prime\prime}$ and following the previously known dusty spiral structures \citep[e.g.][]{davies14}. They describe the CO kinematics as presenting two components, one due to gas rotation and another due to the interaction of the AGN outflow with the molecular gas. In additon, they derived a total molecular gas mass of 1.1$\times$10$^{7}$\,M$_\odot$ for the nuclear disk.

This paper is organized as follows. Section 2 presents the observations and data treatment, Section 3 shows the two-dimensional maps for the emission-line flux distribution and kinematics, as well as for the stellar kinematics. These results are discussed in Section 4 and the conclusions of this work are presented in Section 5.

\section{Observations and Data Reduction}

We use Gemini Observatory archival data of NGC\,5643 obtained with the Gemini Multi-Object Spectrograph Integral Field Unit \citep[GMOS IFU;][]{as02,hook04} on the Gemini South Telescope.  The observations were done in the two-slit mode, resulting in a Field of View (FoV) of 5\arc$\times$7\arc, centered at the nucleus of the galaxy. The R400 grating in combination to the G5325 filter, resulting in a spectral range 7750 -- 9950 \AA. The total on source exposure time was 65\,min, divided in three individual exposures of 22 \,min each.

The data reduction followed the standard procedures \citep[e.g.][]{lena14,brum17} using the {\sc gemini}  package in {\sc iraf} software \citep{tody86,tody93}. These procedures include   trimming of the images, bias subtraction, flat-fielding, cosmic rays cleaning, extraction of the spectra, wavelength calibration using as reference the observed spectra of Ar lamps and sky subtraction. The flux calibration is performed  using a sensitivity
function generated from the spectrum of a photometric standard.  Individual datacubes for each exposure were created at an angular sampling of 0.05\arc$\times$0.05\arc. These datacubes were then median combined using a sigma clipping algorithm to remove spurious features and the location of the continuum peak was used to perform the alignment among the cubes.  

The final datacube of NGC\,5643 covers the inner 5\arc$\times$7\arc\ (285$\times$400~pc$^2$) at an angular resolution of 0.8\arc\ ($\sim$45\,pc), as obtained from measurement of the Full Width at Half Maximum (FWHM) of the flux distribution of the standard star. The velocity resolution is  $\sim$95\,\kms, as the FWHM of typical arc lamp lines used to the wavelength calibration of the spectra.


\section{Results}


\begin{figure*}[h!]
\begin{center}
\includegraphics[width=0.95\textwidth]{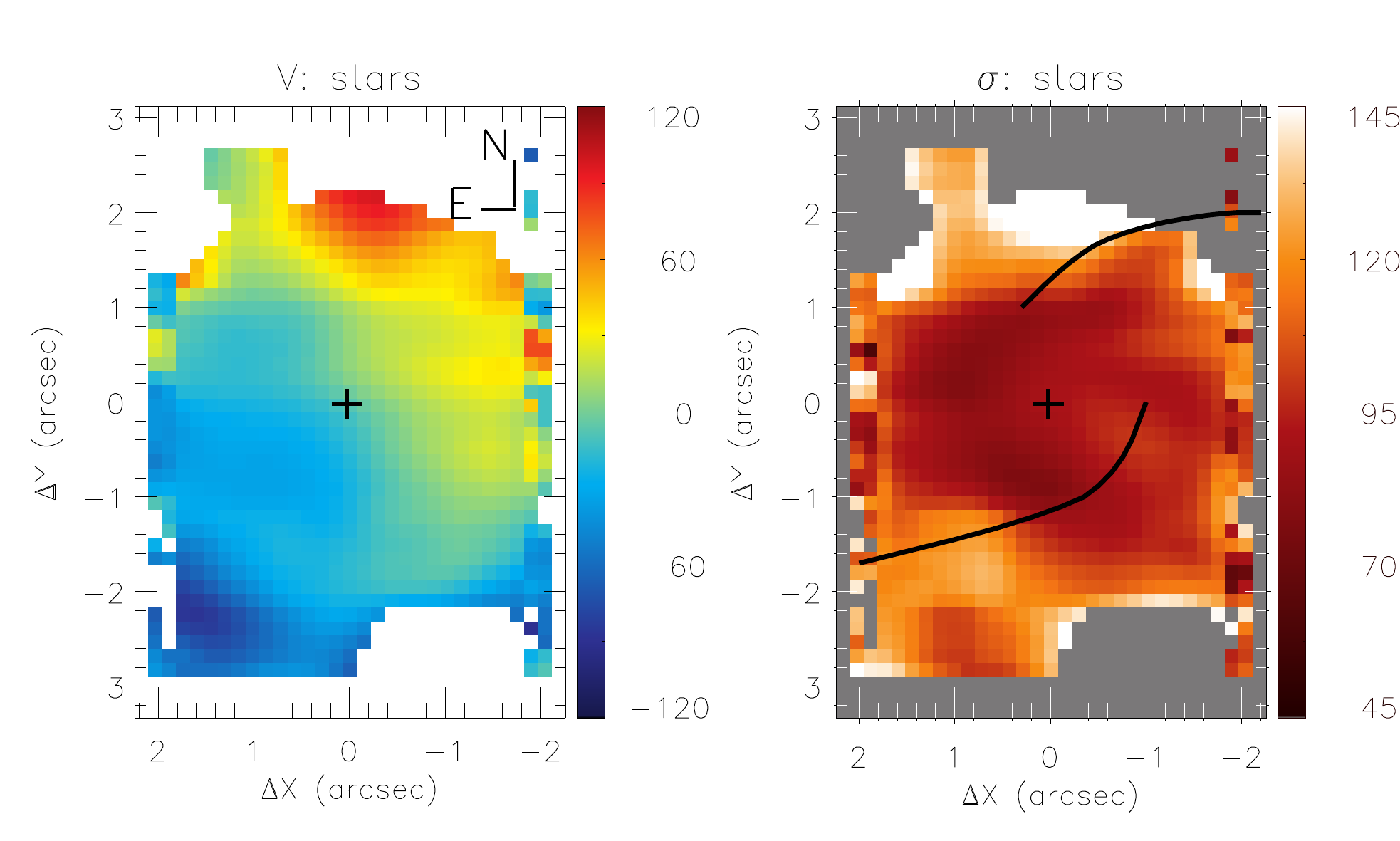}
\caption{Stellar velocity field (left) and stellar velocity dispersion map (right). The white/grey regions in the velocity/$\sigma_\star$ map are masked locations due to the low SNR of the spectra. The color bars show the velocities in \kms\ units and the central cross marks the position of the continuum peak. The continuous lines shown in the $\sigma_\star$ map mark the location of the main dust strucutures as seen in the V-H color map of \citet{davies14}} \label{stel}
\end{center}
\end{figure*}

Figure~\ref{large} (on the top) shows a large scale J-band image of NGC\,5643 obtained from the Two Micron All Sky Survey \citep[2MASS;][]{2mass}. This image clearly reveals the presence of a bar oriented along the east-west direction. The top right panel of Fig.~\ref{large} shows the continuum image obtained from the GMOS datacube, by averaging the fluxes within a spectral window of 300\,\AA\ centered at 8500\,\AA. One can observe that the GMOS flux contours are slightly more elongated along the orientation of the bar. 
The bottom panels show the integrated spectra within apertures of 0.7\arc\ diameter centered at the nucleus (labeled as N in the top-right panel) and at 1.5\arc\ east of it (labeled as A). 
The \siii$\lambda 9069$ emission line and Ca\,{\sc ii}$\lambda\lambda\lambda$8498,8542,8662 absorption triplet are identified in the nuclear spectrum. These features are used to map the gas kinematics and distribution and the stellar kinematics. 

\subsection{Stellar Kinematics}

In order to obtain measurements of the line-of-sight velocity (V$_{\rm los\star}$)  and velocity dispersion ($\sigma_\star$) of the stars in NGC\,5643, we used the  penalized pixel-fitting ({\sc ppxf}) method  \citep{cappellari04,cappellari17} to fit the Ca\,{\sc ii}$\lambda\lambda\lambda$8498,8542,8662 absorption triplet present in the galaxy spectra. As spectral templates, we used selected spectra from the stellar library of \citet{cenarro01}, which cover the spectral range of 8348-9020\AA\ at  spectral resolution of 1.5\,\AA. The choice of this spectral library was done because  it spans  a wide range in stellar atmospheric parameters and the spectral resolution of the spectra is similar to that of our GMOS data.

Before fitting the observed spectra, we have rebinned the datacube to 0.15\arc$\times$0.15\arc\  spaxels in order to increase the signal-to-noise ratio (SNR) and allow reliable measurements.  In Figure~\ref{stel} we show the V$_{\rm los\star}$ (left) and $\sigma_\star$ (right) maps for NGC\,5643. The systemic velocity of the galaxy (1241\,\kms, as derived in Sec.~\ref{disc_stel}) was subtracted from the observed velocity field and the white/grey regions in the  V$_{\rm los\star}$/$\sigma_\star$ maps represent locations where the SNR was not high enough to obtain reliable fits of the observed spectra. At these locations, the uncertainties in V$_{\rm los\star}$ and $\sigma_\star$ are higher than 30\,\kms.  

\begin{figure}[h!]
\begin{center}
\includegraphics[width=0.45\textwidth]{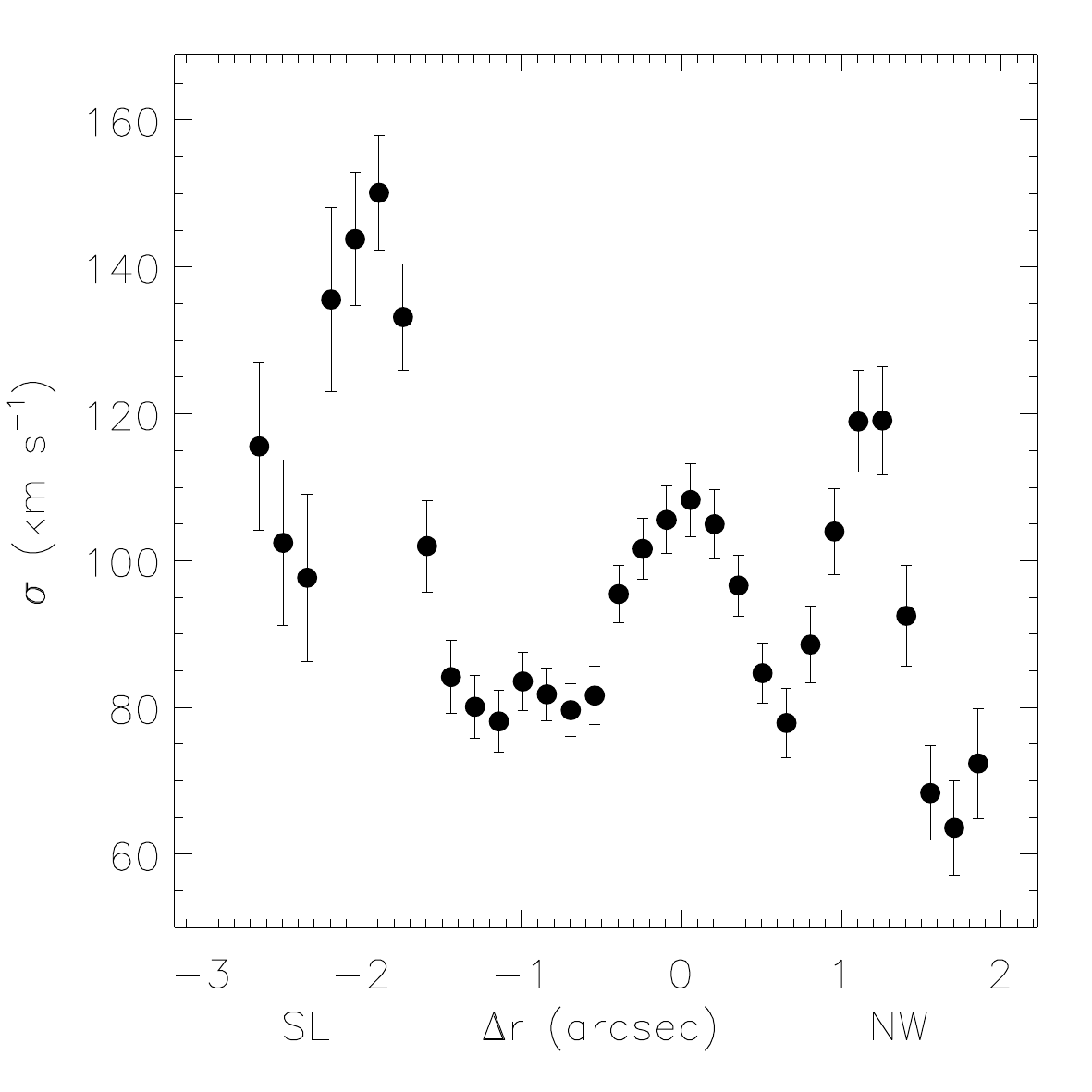}
\caption{One dimensional cut in the $\sigma_\star$ map along the major axis of the galaxy ($\psi_0$=$-36^\circ$). \label{cut}}
\end{center}
\end{figure}

The stellar velocity field (left panel of Fig.\,\ref{stel}) of NGC\,5643 shows a projected velocity amplitude of $\sim$100\,\kms, with redshifts observed to the northwest and blueshifts to the southeast. 
The stellar velocity dispersion map (right panel of Fig.\,\ref{stel}) shows values ranging from 50 to 140~\kms, with the highest values seen mostly to the north and southeast of the nucleus. Surrounding the nucleus, the lowest $\sigma_\star$ are observed, and seems to delineate a partial ring with radius $\sim$1\arc. In Fig.~\ref{cut} we present an one dimensional cut along the major axis of the galaxy ($\psi_0$=$-36^\circ$), extracted within a pseudo slit of 0.45\arc\ width. This plot clearly shows the ring of lower $\sigma_\star$ values. At the nucleus, the $\sigma_\star$ values are around 110\,\kms, then they decrease to $\sigma_\star\sim80$\,\kms\ at the ring region and increase again at larger distances.

\subsection{\siii$\lambda 9069$ flux distribution and  kinematics} 

\begin{figure}[h!]
\begin{center}
\includegraphics[width=0.49\textwidth]{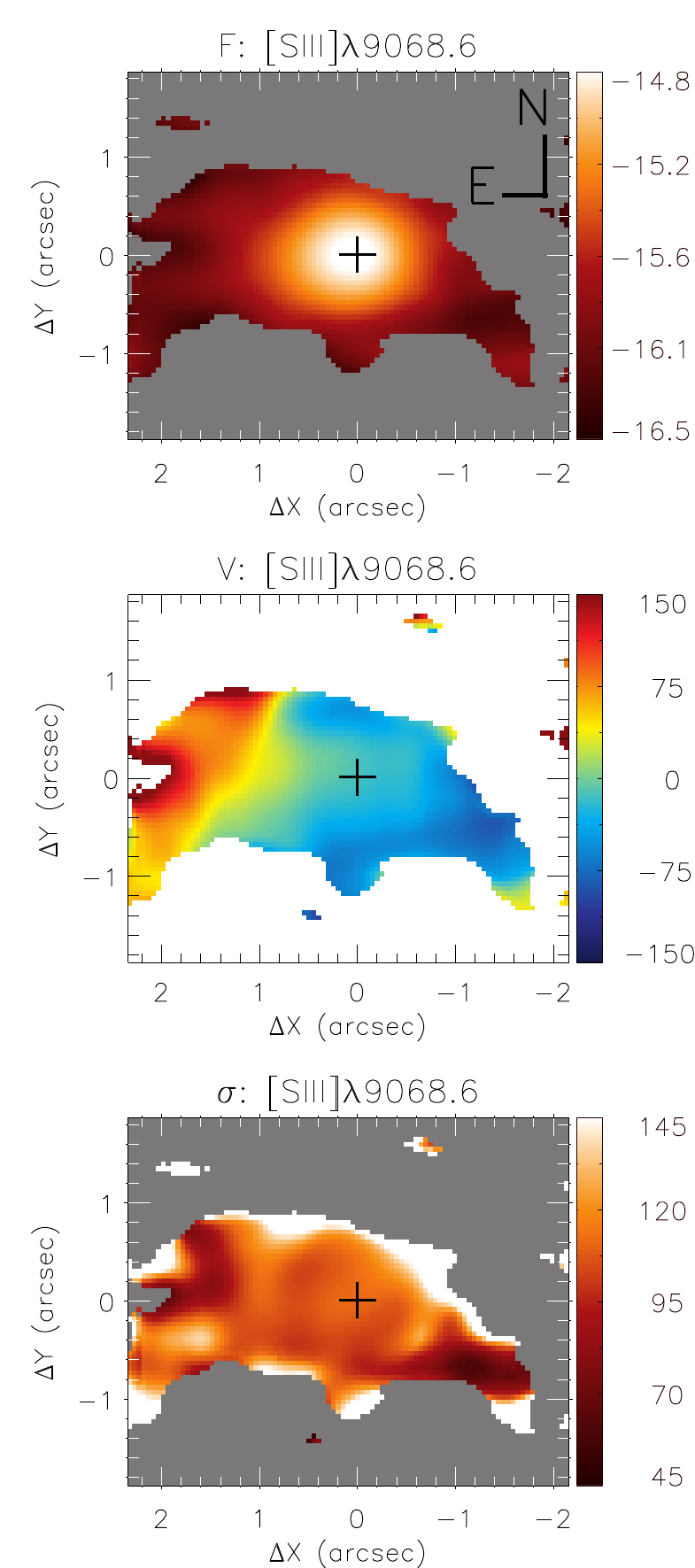}
\caption{\siii$\lambda9069$ flux map (top), velocity field, after the subtraction of the systemic velocity (middle) and velocity dispersion map (bottom). White and grey regions correspond to masked locations due to poor fits or non detection of the line. The fluxes are shown in logarithmic units of  erg\,s$^{-1}$\,cm$^{-2}$\,spaxel$^{-1}$ and the velocities are shown in \kms. The crosses mark the location of the continuum peak. \label{siii}}
\end{center}
\end{figure}

We used the Emission-line PROfile FITting ({\sc profit}) routine \citep{profit} to fit the observed \siii$\lambda 9069$ profile at each spaxel by Gaussian curves and obtain measurements for its flux, centroid velocity (V$_{\rm SIII}$)  and velocity dispersion ($\sigma_{\rm SIII}$). Figure~\ref{siii} shows the corresponding maps. The masked locations correspond to regions where the SNR of the \siii$\lambda 9069$ line was not high enough to obtain good fits. In these maps, we excluded  regions farther than 1.8\arc\ to the north and south directions, where no line emission was detected.

The top panel of Fig.~\ref{siii} presents the \siii$\lambda9069$ flux distribution, which shows an elongated structure along the east-west direction. This structure extends to up to 2.2\arc\ to the east  of the nucleus and is narrower to the west side of it, extending to 1.8\arc. Along the north-south direction, the \siii\ shows extended emission only at locations closer than 1\arc\ of the nucleus of the galaxy. 

The \siii\ velocity field is shown in the middle panel of Fig.~\ref{siii}, after the subtraction of the systemic velocity of the galaxy, as obtained by the fitting of the stellar velocity field by a rotation disk model (see Sec. \ref{disc_stel}). The highest redshifts of up to 150\,\kms\ are seen east of the nucleus, while similar velocities in blueshifts are observed west of the nucleus.

The bottom panel of Fig.~\ref{siii} shows the resulting $\sigma_{\rm SIII}$ map. The $\sigma_{\rm SIII}$ values were corrected for the instrumental broadening and range from 40 to 150~\kms. At most locations $\sigma_{\rm SIII}\approx90-120$\,\kms.

\begin{figure*}[h!]
\begin{center}
\includegraphics[width=0.95\textwidth]{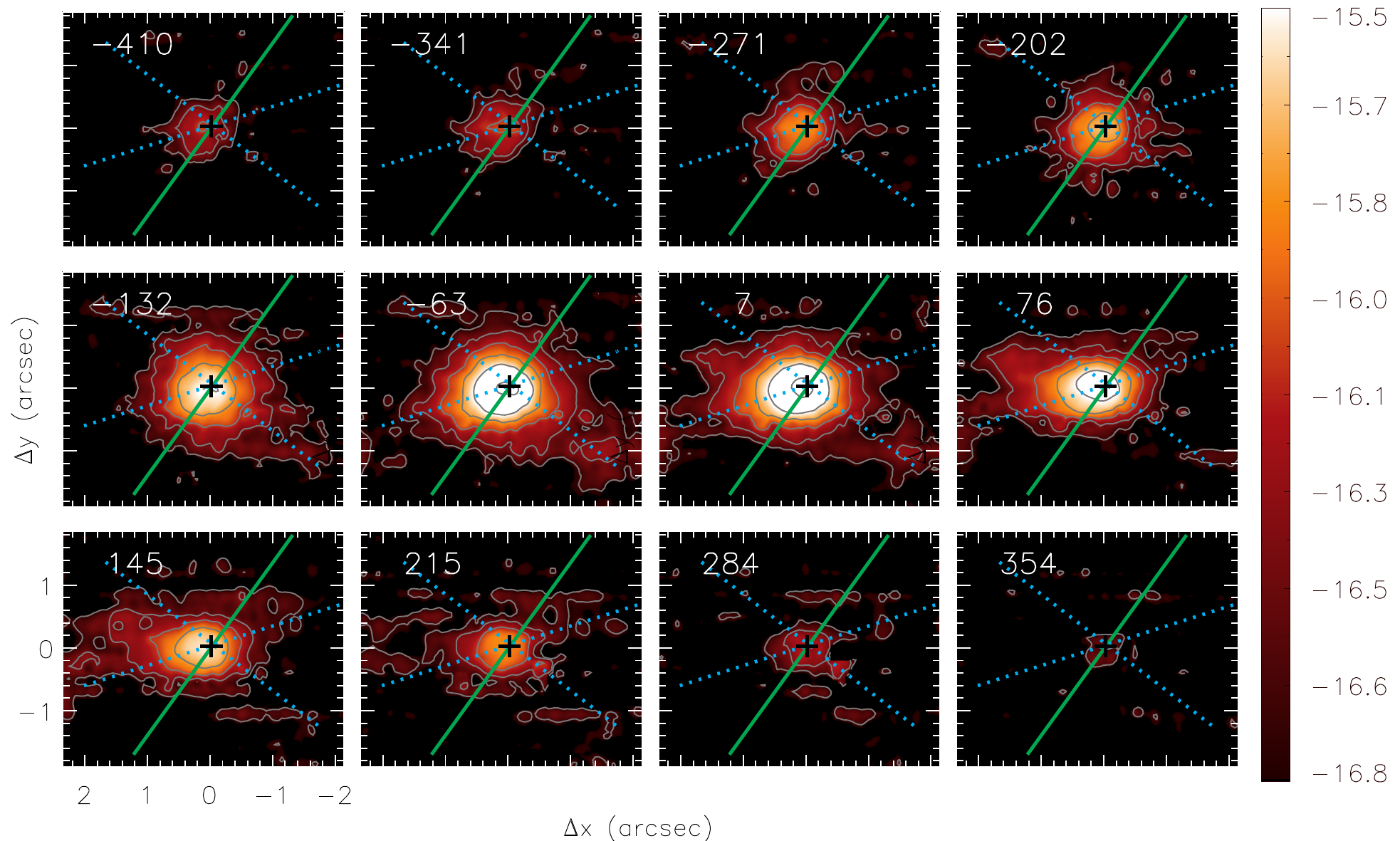}
\caption{Velocity channel maps along the \siii$\lambda9069$ emission-line profile for a velocity bin of $\sim$70\,\kms\ (3 pixels). The color bar shows the fluxes logarithmic  units of  erg\,s$^{-1}$\,cm$^{-2}$\,spaxel$^{-1}$. The velocities relative to the systemic velocity of the galaxy are shown in the top-left corner of each panel. The continuous green lines represent the orientation of the line of nodes, as derived from the stellar velocity field. The dotted lines delineate the walls of the bi-cone, derived by \citet{fischer13}. The central crosses mark the location of the continuum peak. } \label{cham}
\end{center}
\end{figure*}

The relatively high spectral resolution of the GMOS spectra has allowed us to slice the \siii\
emission-line profile into a sequence of velocity bins and construct the velocity channel maps,  shown in Figure~\ref{cham}. These maps allow a better sample of the gas kinematics over the whole velocity distribution, including the wings of the line profiles.  The velocity slices within bins of $\sim$70\,\kms\ (corresponding to three spectral pixels) were obtained after  subtraction of the continuum, determined as averages of the fluxes from both sides of the emission line.  Each panel presents flux levels in logarithmic units for the velocity slice shown. The zero velocity is adopted as the value obtained from the modeling of the stellar velocity field (Sec.~\ref{disc_stel}).  The continuous line shown  in each panel of Fig.~\ref{cham} represents the orientation of the line-of-nodes of the galaxy, the dotted lines delineate the bi-cone geometry, as obtained by \citet{fischer13} and the central cross marks the location of the nucleus. 

The channel maps trace the gas from  negative velocities (blueshifts) to  positive values (redshifts) relative to the systemic velocity of the galaxy. The highest blueshifts (of up to $-410$\,\kms) and redshifts (of up to 350\,\kms)  are observed mainly at the nucleus of the galaxy. For smaller blueshifts, besides the nuclear emission, an elongated structure to the west-southwest is observed, co-spatial with the stripe seen in the \siii\ flux map (top panel of Fig.~\ref{siii}). A similar redshifted structure is observed to the east. In addition, some  emission northwest of the nucleus is seen in redshift, following the orientation of the major axis of the galaxy (green line), clearly seen at channel maps centered in velocities in the range 76--215\,\kms. A slightly elongated structure along the major axis of the galaxy is also seen in blueshifts to the southeast, clearly observed in panels centered at $-341$ and $-271$\,\kms.

\section{Discussions}

\subsection{Stellar Kinematics}\label{disc_stel}

The stellar velocity field  (Fig.~\ref{stel}) of NGC\,5643  shows a clear rotation pattern with the northwest side of the galaxy receding and the southeast side approaching, consistent with the velocity field presented by \citet{davies14}, as obtained from the fit of the CO\,$\lambda2.3\,\mu$m bands using VLT SINFONI observations. In addition, a similar behavior is seen in the velocity field for the cold molecular gas as obtained from ALMA $^{12}$CO(2--1) line observations by \citet{ah18}, but the CO presents smaller velocity dispersion values than that of the stars, consistent with the fact that the cold molecular gas being located on a thin disk, whereas the bulge stars contributes to the observed stellar kinematics. 

We fitted the observed $V_{\rm los\star}$ by an analytical model, under the assumption that the stars  moves in circular orbits in the plane of the galaxy, within a central gravitational potential \citep{bertola91}. In this model, the rotation velocity field is given by:

\[ 
 V_{mod}(R,\psi)=V_{s}+ 
\]
\begin{equation}
    \frac{AR\cos(\psi-\psi_{0})\sin(i){\cos^{p}(i)}}{\{R^{2}[\sin^{2}(\psi-\psi_{0})+\cos^{2}(i)\cos^{2}(\psi-\psi_{0})]+{c_{0}}^{2}\cos^{2}(i)\}^{\frac{p}{2}}},
    \end{equation}
where $R$ and $\psi$ are the coordinates of each spaxel in the plane of the sky, $V_{s}$ is the systemic velocity of the galaxy, $A$ is the velocity amplitude, $\psi_{0}$ is the position angle of the major axis, $i$ is the disc inclination relative to the plane of the sky. 
The {\it p} parameter measures the slope of the rotation curve where it flattens, being limited between 1 $\le$ {\it p} $\le$ 3/2. For {\it p}\,=\,1 the rotation curve at large radii is asymptotically flat
while for {\it p}\,=\,3/2 the system has a finite mass. $c_{0}$ is a concentration parameter,  defined as the radius where the rotation curve reaches 70\,\% of its velocity amplitude.

 We fitted the model to the observed velocities using the {\sc mpfitfun} routine \citep{markwardt09} to perform a least-squares fit, in which initial guesses are given for the free parameters. As the GMOS FoV is small, the position of the kinematical center was kept fixed to the location of the continuum peak and the disc inclination was fixed to $i=34^\circ$ \citep{davies14}.
 
\begin{figure*}[h!]
\begin{center}
\includegraphics[width=0.95\textwidth]{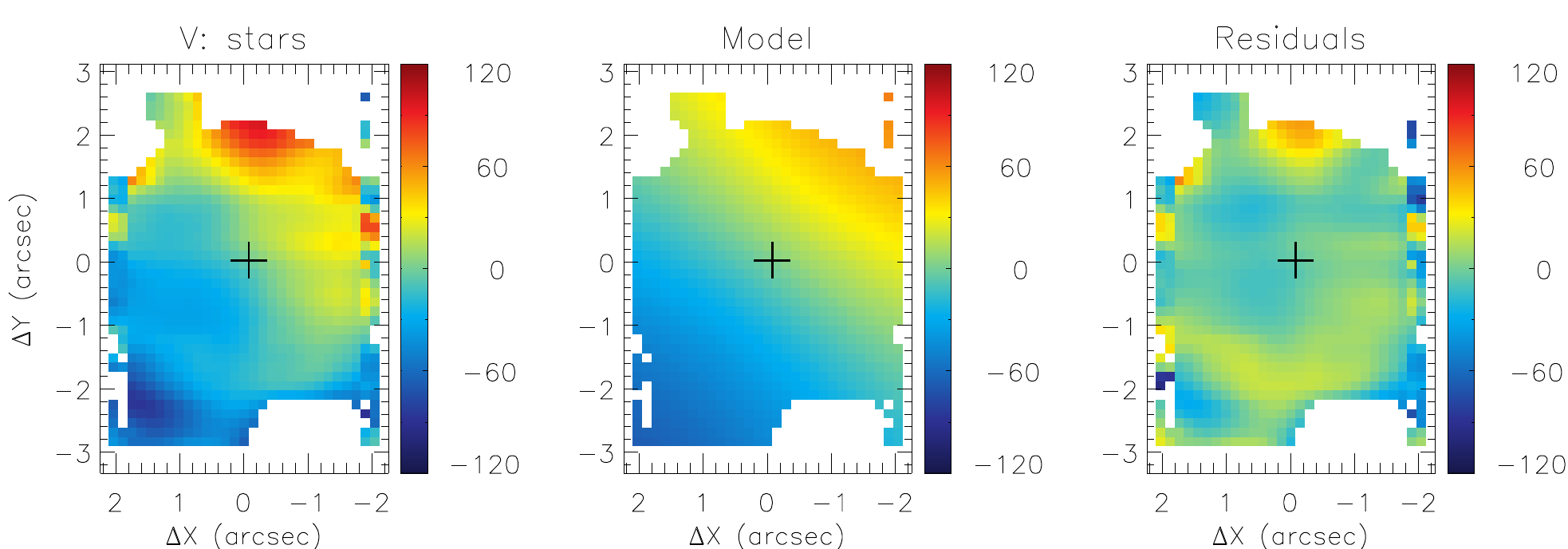}
\caption{Left: stellar velocity field: Middle: rotating disc model for the stellar velocity field. Right: residual map between the observed and modeled velocities. The color bar shows the range of velocities in km\,s$^{-1}$ and the cross marks 
the position of the nucleus.  \label{bertola}}
\end{center}
\end{figure*}

Figure~\ref{bertola} shows the stellar velocity field (left panel), the resulting best fit model (middle) and a residual map (right), obtained by subtracting the model from the observed velocity field.  The residuals are smaller than 30~\kms at all locations, indicating that the observed velocities are well reproduced by the  model. 

The resulting parameters for the best fit model are: $V_s=1241\pm7$\,\kms, relative to the heliocentric rest frame, $A=172\,\pm22$\,\kms, $\psi_0$=$-36^\circ\pm$ 3$^\circ$, $p=1$ and $c_0=3.7^{\prime\prime}\pm0.6^{\prime\prime}$.  
The systemic velocity is about 40\,\kms\ larger than the one obtained from 21 cm H\,{\sc i} line \citep[$1199\pm2$\,\kms][]{systevel}, possible due to the very distinct apertures used in the H\,{\sc i} and GMOS observations. The orientation of the line of nodes is consistent with the value presented by \citet{davies14} derived from the stellar kinematics measured by fitting the CO absorption bandheads in the K-band.

The velocity residual map shows small values at all locations, with a mean absolute value of $\sim$10~\kms, and the deprojected velocity amplitude is larger than the $\sigma_\star$ values, indicating that the stellar kinematics in the inner region of NGC\,5643 is dominated by regular rotation. This result can be compared with previous studies for neaby galaxies. In \citet{Riffel_LLP_stel}, we presented stellar kinematics measurements for 16 nearby Seyfert galaxies, derived by the fitting of the CO absorption bandheads in the K-band using Gemini NIFS observations.  We found that the stellar kinematics in the inner few of hundred of parsecs is dominated by a disk-like component and the stellar velocity fields are well reproduced by a rotating disk model, with kinematic axes that follows the same orientation of large scale disk. \citet{dumas07} used optical IFS to map the stellar kinematics of a sample of 39 active galaxies and a matched control sample of inactive galaxies, selected to have similar blue magnitudes, Hubble type and inclinations. They found that for both active and inactive galaxies, the stellar kinematics in the central region is dominated by a disk component. Similar results were also found by other authors for neaby galaxies \citep{barbosa06,fb06}. Thus, our results for NGC\,5643 are in agreement with previous studies and suggest that the motion of the stars in the inner 200~pc is dominated by circular orbits at the plane of the galaxy due the gravitational potential of the bulge, as the FoV of our observations is smaller than the bulge length of NGC\,5643.

The stellar velocity dispersion map (right panel of Fig.~\ref{stel}) shows values in the range 50--140~\kms. This range is similar to that derived using the CO absorptions at 2.3\,$\mu$m  presented by \citet{hicks13}. However, their map is much noisier than ours, which suggests in addition a ring of low-$\sigma_\star$ values ($\sim$70--80 \kms) surrounding the nucleus (for which $\sigma_\star\sim100$\,\kms) at a distance of 0.8\arc\ ($\sim$50 pc). Similar rings have been observed for other Seyfert galaxies at scales of a few hundred parsecs and attributed to young/intermediate age stellar populations \citep[e.g.][]{mrk1066:pop,mrk1157:pop,Riffel_LLP_stel,diniz17}. We do not observe any clear correlation between the structures seen in the $\sigma_\star$ map and the known dust strucutures \citep{martini03,davies14}, as indicated by the lines drawn on the $\sigma_\star$ map (Fig.~\ref{stel}), which trace the main dusty structures seen in the V-H color map of \citet{davies14}.

\subsection{The \siii\ Emission and Kinematics}

 The HST \citep{simpson97,fischer13} and VLT MUSE \citep{cresci15} [O\,{\sc iii}]$\lambda5007$  flux distributions for 
NGC\,5643 reveal a well defined triangular emission region east of the nucleus, extending to up to 1.8~kpc and showing  several knots of emission. The collimated structure is seen also in high resolution flux maps for the Br$\gamma$ \citep[from VLT SINFONI observations -- ][]{davies14,menezes15} and for the H$\alpha$ \citep[from VLT MUSE observations -- ][]{cresci15} emission lines. Besides the emission region east of the nucleus, the  Br$\gamma$ and H$\alpha$ maps clearly show extended emission to the west--southwest of the nucleus.  Our  \siii$\lambda9069$ flux map (Fig.~\ref{siii}) is consistent with the [O\,{\sc iii}]$\lambda5007$ and Br$\gamma$ flux maps, considering that the GMOS angular resolution is worse than that of HST and VLT--SINFONI.

The \siii\ velocity field (Fig.~\ref{siii}) is similar to that of Br$\gamma$ \citep{davies14,menezes15}, showing redshifts to the east of the nucleus and blueshifts to the west.  The \siii\ kinematics is also consistent with the orientation of the outflows observed in [O\,{\sc iii}] by \citet{cresci15}.  
The [O\,{\sc iii}] kinematics is modeled by \citet{fischer13} as a bi-cone with axis oriented along $PA=80^\circ$, being displaced by  $i_{\rm cone}=65^\circ$ from the line-of-sight and with inner and outer opening angles of $\theta_{\rm min}=50^\circ$ and $\theta_{\rm max}=55^\circ$, respectively. \citet{davies14} argue that the Br$\gamma$ and H$_2$ 1--0 S(1) kinematics are  consistent with this model. Besides the \siii\ velocity field, the velocity channel maps  are consistent with outflows within a bi-cone, as most of the redshifts and blueshifts are seen within the doted lines in Fig.~\ref{cham}, that delineates the geometry of the bi-cone model of \citet{fischer13}.
Thus, we conclude that the \siii\ emission  is originated from the same outflowing gas that originates the [O\,{\sc iii}] and Br$\gamma$ emission.  This interpretation is further supported by the CO velocity residual map presented by \citet{ah18}, that shows redshifts to the east and blueshifts to the west of the nucleus, being interpreted by the authors as due to radial movements of material  being pushed  outwards in the galaxy disk by the ionized gas outflow.

Aside from the outflowing gas component, the \siii\ velocity channel maps (Fig.~\ref{cham}) show  a structure in redshifts northwest of the nucleus, extending to up to 1.5\arc\  (clearly observed in velocity slices from 76--215\,\kms). Another slightly elongated structure is seen in blueshits to the southeast (clearly seen in panels centered at $-341$ and $-271$\,\kms). These structures are seen at the same orientation of the kinematic major axis of the galaxy,  as observed in the stellar  (Fig.~\ref{stel}) and CO \citep{ah18} velocity fields, that shows redshifts to the northwest and blueshifts to the southeast. The most plausible interpretation of these \siii\ kinematic components is that they are due to emission of gas at the plane of the disc, possible heated by the outflowing gas and AGN radiation, as the bi-cone intercepts the disc plane.

Considering that the bulk of the \siii\ kinematics is consistent with outflows within a bi-cone, we can use the geometric parameters of the bi-cone, in combination with our velocity measurements to estimate the  mass outflow rate in ionized gas. Assuming that the redshifts seen to the east are originated from gas located in the front wall of the bi-cone, the observed projected velocity of $\sim$100~\kms\ (from Fig.~\ref{siii})  corresponds to outflows with velocity of $v_{\rm out}\approx 100/{\rm sin}(i_{\rm cone}-(\theta_{\rm max}-\theta_{\rm min})/2) \approx 330$\,\kms. We used the geometric parameters of the bi-cone from  \citet{fischer13} and we note that $v_{\rm out}$ is consistent with their model (their Fig.~22). The ionized gas mass outflow rate can be estimated by
\begin{equation}
\dot{M}=2\,m_p\,N_e\,A\, v_{\rm out}\, f. 
\end{equation}
The factor 2 is included to consider both sides of the bi-cone, $m_p$ is the proton mass, $N_e$ is the electron density, $f$ is the filling factor,  $A=\pi r^2\,(\tan\,\theta_{\rm max}-\tan\,\theta_{\rm min})$ is the cross section of the outflow, and $r$ is the distance from the nucleus along the bi-cone axis. Assuming typical values of $f=0.01$ and $N_e=500$\,cm$^{-3}$ \citep[e.g.][]{osterbrock06,dors15} and $r=$1\arc$\approx50\,$pc, we obtain $\dot{M}\approx\,0.3\,$\,M$_\odot$\,yr$^{-1}$ for NGC\,5643. 

The mass outflow-rate derived for NGC\,5643 is within the range of values observed in other Seyfert galaxies \citep[$10^{-2}-10^{1}$\,M$_\odot$\,yr$^{-1}$; ][]{barbosa09,baaa13,revalski18}. 
 This value can be compared to the accretion rate necessary to power the AGN at the nucleus of NGC\,5643, which can be derived by
\begin{equation}
 \dot{m}=\frac{L_{\rm bol}}{c^2\eta},
\end{equation}
where $L_{\rm bol}$ is the nuclear bolometric luminosity, $\eta$ is the efficiency 
of conversion of the rest mass energy of the accreted material into radiation and $c$ 
is the light speed. Using $L_{\rm bol}\approx1\times10^{43}$ erg\,s$^{-1}$ from \citet{Brightman17} and  assuming $\eta\approx0.1$, which is a typical value for a ``standard'' geometrically thin, optically thick accretion disc \citep[e.g.][]{frank02}, we obtain a mass accretion rate of $\dot{m}\approx1.7\times10^{-3}~{\rm M_\odot\, yr^{-1}}$.

The mass outflow rate ($\dot{M}$) estimated for NGC\,5643 is about two orders of magnitude larger than $\dot{m}$. This result is consistent with those obtained for other Seyfert galaxies \citep[e.g.][]{baaa13} and  indicates that most of the outflowing gas observed in the NLR of NGC\,5643 does not originate in the AGN, but is a result of the interaction of winds launched by the accretion disk and the ambient gas, which is pushed away by the nuclear outflow.  

Finally, we can use the derived mass-outflow rate for NGC\,5643 to verify if it follows the same AGN wind scaling relation between $\dot{M}$ and the bolometric luminosity (L$_{\rm bol}$) observed for a sample of 94 AGN \citep{fiore}. Considering that NGC\,5643 presents log\,L$_{\rm bol}=43\pm0.5$ \citep{Brightman17} and using the value of   $\dot{M}\approx\,0.3\,$\,M$_\odot$\,yr$^{-1}$ derived above, we note that NGC\,5643 lies at the lower end of observed correlation between $\dot{M}$ and L$_{\rm bol}$ in ionized gas \citep[see Fig.\,1 of ][]{fiore}, suggesting that the observed correlation can be extended to lower luminosity AGN.

\section{Conclusions}

We used GMOS IFU observations of the inner 285$\times$400\,pc$^2$ of the Seyfert 2 galaxy NGC\,5643 to map the gas and stellar kinematics at a velocity resolution of 95\,\kms\ (FWHM) and spatial resolution of 45~pc. Our main conclusions are:

\begin{itemize}
\item The \siii$\lambda9069$ flux  map shows a triangular emission region east of the nucleus, extending to up to the border of the field of view (140\,pc). A more collimated structure is seen to the west-southwest to up to 100\,pc from the nucleus. The \siii$\lambda9069$ flux distribution is similar to that [O\,{\sc iii}]$\lambda5007$ and H\,{\sc i} recombination lines, previously published.

\item The \siii\ kinematics is dominated by outflows within a bi-cone oriented along $PA\approx80^{\circ}$. The projected velocity of the outflow is about 100\,\kms\ and we derive a mass outflow rate of 0.3\,M$_\odot$\,yr$^{-1}$ in ionized gas. 

\item Velocity channel maps along the \siii$\lambda9069$ emission-line profile reveal a secondary kinematic component, originated from gas located in the plane of the galaxy, possible heated by the outflowing material.

\item The stellar velocity field shows regular rotation with a projected velocity amplitude of 100\,\kms\ and is well reproduced by a model of a rotating disc with major axis oriented along $\psi_0=-36^0$. 

\item The stellar velocity dispersion map shows values ranging from 50 to 140~\kms\ and suggests the presence of a ring of low-$\sigma_\star$ values (80~\kms) surrounding the nucleus at a distance of 50 pc from it. This structure is attributed to young/intermediate age stellar populations, which still preserve the kinematics of the gas whence they were formed.  

\end{itemize}

\begin{acknowledgements}
We thank the two anonymous referees for valuable suggestions which helped to improve the paper. This work has been partially supported by the Conselho Nacional de Desenvolvimento Cient\'{i}fico e Tecnol\'{o}gico (CNPq), Coordena\c{c}\~{a}o de Aperfei\c{c}oamento de Pessoal de N\'{i}vel Superior (CAPES) and Funda\c c\~ao de Amparo \`a Pesquisa do Estado do Rio Grande do Sul (FAPERGS).
Based on observations obtained at the Gemini Observatory, acquired through the Gemini Observatory Archive and processed using the Gemini IRAF package, which is operated by the Association of Universities for Research in Astronomy, Inc., under a cooperative agreement with the NSF on behalf of the Gemini partnership: the National Science Foundation (United States), the National Research Council (Canada), CONICYT (Chile), Ministerio de Ciencia, Tecnolog\'{i}a e Innovaci\'{o}n Productiva (Argentina), and Minist\'{e}rio da Ci\^{e}ncia, Tecnologia e Inova\c{c}\~{a}o (Brazil).
\end{acknowledgements}

\bibliography{ngc5643.bbl}

\begin{thebibliography}{}
\makeatletter
\relax
\def\mn@urlcharsother{\let\do\@makeother \do\$\do\&\do\#\do\^\do\_\do\%\do\~}
\definecolor{darkblue}{rgb}{0,0,0.597656}
\def\mndoi{\begingroup\mn@urlcharsother \@ifnextchar [ {\mndoi@} {\mndoi@[]}}
\def\mndoi@[#1]#2{\def\@tempa{#1}\ifx\@tempa\@empty \href
  {http://dx.doi.org/#2} {\textcolor{darkblue}{doi:#2}}\else \href
  {http://dx.doi.org/#2} {\textcolor{darkblue}{#1}}\fi \endgroup}
\def\mn@eprint#1#2{\mn@eprint@#1:#2::\@nil}
\def\mn@eprint@arXiv#1{\href {http://arxiv.org/abs/#1} {{\tt arXiv:#1}}}
\def\mn@eprint@dblp#1{\href {http://dblp.uni-trier.de/rec/bibtex/#1.xml}
  {dblp:#1}}
\def\mn@eprint@#1:#2:#3:#4\@nil{\def\@tempa {#1}\def\@tempb {#2}\def\@tempc
  {#3}\ifx \@tempc \@empty \let \@tempc \@tempb \let \@tempb \@tempa \fi \ifx
  \@tempb \@empty \def\@tempb {arXiv}\fi \@ifundefined
  {mn@eprint@\@tempb}{\@tempb:\@tempc}{\expandafter \expandafter \csname
  mn@eprint@\@tempb\endcsname \expandafter{\@tempc}}}

\bibitem[\protect\citeauthoryear{{Allington-Smith} et~al.,}{{Allington-Smith}
  et~al.}{2002}]{as02}
{Allington-Smith} J.,  et~al., 2002, \mndoi [\pasp] {10.1086/341712}, \href
  {http://adsabs.harvard.edu/abs/2002PASP..114..892A} {114, 892}

\bibitem[\protect\citeauthoryear{{Alonso-Herrero} et~al.,}{{Alonso-Herrero}
  et~al.}{2018}]{ah18}
{Alonso-Herrero} A.,  et~al., 2018, \mndoi [\apj] {10.3847/1538-4357/aabe30},
  \href {http://adsabs.harvard.edu/abs/2018ApJ...859..144A} {859, 144}

\bibitem[\protect\citeauthoryear{{Antonucci}}{{Antonucci}}{1993}]{antonucci93}
{Antonucci} R.,  1993, \mndoi [\araa] {10.1146/annurev.aa.31.090193.002353},
  \href {http://adsabs.harvard.edu/abs/1993ARA%26A..31..473A} {31, 473}

\bibitem[\protect\citeauthoryear{{Barbosa}, {Storchi-Bergmann}, {Cid
  Fernandes}, {Winge}  \& {Schmitt}}{{Barbosa} et~al.}{2006}]{barbosa06}
{Barbosa} F.~K.~B.,  {Storchi-Bergmann} T.,  {Cid Fernandes} R.,  {Winge} C.,
  {Schmitt} H.,  2006, \mndoi [\mnras] {10.1111/j.1365-2966.2006.10690.x},
  \href {http://adsabs.harvard.edu/abs/2006MNRAS.371..170B} {371, 170}

\bibitem[\protect\citeauthoryear{{Barbosa}, {Storchi-Bergmann}, {Cid
  Fernandes}, {Winge}  \& {Schmitt}}{{Barbosa} et~al.}{2009}]{barbosa09}
{Barbosa} F.~K.~B.,  {Storchi-Bergmann} T.,  {Cid Fernandes} R.,  {Winge} C.,
  {Schmitt} H.,  2009, \mndoi [\mnras] {10.1111/j.1365-2966.2009.14485.x},
  \href {http://adsabs.harvard.edu/abs/2009MNRAS.396....2B} {396, 2}

\bibitem[\protect\citeauthoryear{{Bertola}, {Bettoni}, {Danziger}, {Sadler},
  {Sparke}  \& {de Zeeuw}}{{Bertola} et~al.}{1991}]{bertola91}
{Bertola} F.,  {Bettoni} D.,  {Danziger} J.,  {Sadler} E.,  {Sparke} L.,   {de
  Zeeuw} T.,  1991, \mndoi [\apj] {10.1086/170058}, \href
  {http://adsabs.harvard.edu/abs/1991ApJ...373..369B} {373, 369}

\bibitem[\protect\citeauthoryear{{Bianchi}, {Guainazzi}  \&
  {Chiaberge}}{{Bianchi} et~al.}{2006}]{bianchi06}
{Bianchi} S.,  {Guainazzi} M.,   {Chiaberge} M.,  2006, \mndoi [\aap]
  {10.1051/0004-6361:20054091}, \href
  {http://adsabs.harvard.edu/abs/2006A%26A...448..499B} {448, 499}

\bibitem[\protect\citeauthoryear{{Brightman} et~al.,}{{Brightman}
  et~al.}{2017}]{Brightman17}
{Brightman} M.,  et~al., 2017, \mndoi [\apj] {10.3847/1538-4357/aa75c9}, \href
  {http://adsabs.harvard.edu/abs/2017ApJ...844...10B} {844, 10}

\bibitem[\protect\citeauthoryear{{Brum}, {Riffel}, {Storchi-Bergmann},
  {Robinson}, {Schnorr M{\"u}ller}  \& {Lena}}{{Brum} et~al.}{2017}]{brum17}
{Brum} C.,  {Riffel} R.~A.,  {Storchi-Bergmann} T.,  {Robinson} A.,  {Schnorr
  M{\"u}ller} A.,   {Lena} D.,  2017, \mndoi [\mnras] {10.1093/mnras/stx964},
  \href {http://adsabs.harvard.edu/abs/2017MNRAS.469.3405B} {469, 3405}

\bibitem[\protect\citeauthoryear{{Cappellari}}{{Cappellari}}{2017}]{cappellari17}
{Cappellari} M.,  2017, \mndoi [\mnras] {10.1093/mnras/stw3020}, \href
  {http://adsabs.harvard.edu/abs/2017MNRAS.466..798C} {466, 798}

\bibitem[\protect\citeauthoryear{{Cappellari} \& {Emsellem}}{{Cappellari} \&
  {Emsellem}}{2004}]{cappellari04}
{Cappellari} M.,  {Emsellem} E.,  2004, \mndoi [\pasp] {10.1086/381875}, \href
  {http://adsabs.harvard.edu/abs/2004PASP..116..138C} {116, 138}

\bibitem[\protect\citeauthoryear{{Cenarro}, {Cardiel}, {Gorgas}, {Peletier},
  {Vazdekis}  \& {Prada}}{{Cenarro} et~al.}{2001}]{cenarro01}
{Cenarro} A.~J.,  {Cardiel} N.,  {Gorgas} J.,  {Peletier} R.~F.,  {Vazdekis}
  A.,   {Prada} F.,  2001, \mndoi [\mnras] {10.1046/j.1365-8711.2001.04688.x},
  \href {http://adsabs.harvard.edu/abs/2001MNRAS.326..959C} {326, 959}

\bibitem[\protect\citeauthoryear{{Cresci} et~al.,}{{Cresci}
  et~al.}{2015}]{cresci15}
{Cresci} G.,  et~al., 2015, \mndoi [\aap] {10.1051/0004-6361/201526581}, \href
  {http://adsabs.harvard.edu/abs/2015A%26A...582A..63C} {582, A63}

\bibitem[\protect\citeauthoryear{{Davies} et~al.,}{{Davies}
  et~al.}{2014}]{davies14}
{Davies} R.~I.,  et~al., 2014, \mndoi [\apj] {10.1088/0004-637X/792/2/101},
  \href {http://adsabs.harvard.edu/abs/2014ApJ...792..101D} {792, 101}

\bibitem[\protect\citeauthoryear{{Diniz}, {Riffel}, {Riffel}, {Crenshaw},
  {Storchi-Bergmann}, {Fischer}, {Schmitt}  \& {Kraemer}}{{Diniz}
  et~al.}{2017}]{diniz17}
{Diniz} M.~R.,  {Riffel} R.~A.,  {Riffel} R.,  {Crenshaw} D.~M.,
  {Storchi-Bergmann} T.,  {Fischer} T.~C.,  {Schmitt} H.~R.,   {Kraemer} S.~B.,
   2017, \mndoi [\mnras] {10.1093/mnras/stx1006}, \href
  {http://adsabs.harvard.edu/abs/2017MNRAS.469.3286D} {469, 3286}

\bibitem[\protect\citeauthoryear{{Dors}, {Cardaci}, {H{\"a}gele}, {Rodrigues},
  {Grebel}, {Pilyugin}, {Freitas-Lemes}  \& {Krabbe}}{{Dors}
  et~al.}{2015}]{dors15}
{Dors} O.~L.,  {Cardaci} M.~V.,  {H{\"a}gele} G.~F.,  {Rodrigues} I.,  {Grebel}
  E.~K.,  {Pilyugin} L.~S.,  {Freitas-Lemes} P.,   {Krabbe} A.~C.,  2015,
  \mndoi [\mnras] {10.1093/mnras/stv1916}, \href
  {http://adsabs.harvard.edu/abs/2015MNRAS.453.4102D} {453, 4102}

\bibitem[\protect\citeauthoryear{{Dumas}, {Mundell}, {Emsellem}  \&
  {Nagar}}{{Dumas} et~al.}{2007}]{dumas07}
{Dumas} G.,  {Mundell} C.~G.,  {Emsellem} E.,   {Nagar} N.~M.,  2007, \mndoi
  [\mnras] {10.1111/j.1365-2966.2007.12014.x}, \href
  {http://adsabs.harvard.edu/abs/2007MNRAS.379.1249D} {379, 1249}

\bibitem[\protect\citeauthoryear{{Falc{\'o}n-Barroso}
  et~al.,}{{Falc{\'o}n-Barroso} et~al.}{2006}]{fb06}
{Falc{\'o}n-Barroso} J.,  et~al., 2006, \mndoi [\mnras]
  {10.1111/j.1365-2966.2006.10261.x}, \href
  {http://adsabs.harvard.edu/abs/2006MNRAS.369..529F} {369, 529}

\bibitem[\protect\citeauthoryear{{Fiore} et~al.,}{{Fiore} et~al.}{2017}]{fiore}
{Fiore} F.,  et~al., 2017, \mndoi [\aap] {10.1051/0004-6361/201629478}, \href
  {http://adsabs.harvard.edu/abs/2017A%26A...601A.143F} {601, A143}

\bibitem[\protect\citeauthoryear{{Fischer}, {Crenshaw}, {Kraemer}  \&
  {Schmitt}}{{Fischer} et~al.}{2013}]{fischer13}
{Fischer} T.~C.,  {Crenshaw} D.~M.,  {Kraemer} S.~B.,   {Schmitt} H.~R.,  2013,
  \mndoi [\apjs] {10.1088/0067-0049/209/1/1}, \href
  {http://adsabs.harvard.edu/abs/2013ApJS..209....1F} {209, 1}

\bibitem[\protect\citeauthoryear{{Fischer} et~al.,}{{Fischer}
  et~al.}{2017}]{fischer17}
{Fischer} T.~C.,  et~al., 2017, \mndoi [\apj] {10.3847/1538-4357/834/1/30},
  \href {http://adsabs.harvard.edu/abs/2017ApJ...834...30F} {834, 30}

\bibitem[\protect\citeauthoryear{{Frank}, {King}  \& {Raine}}{{Frank}
  et~al.}{2002}]{frank02}
{Frank} J.,  {King} A.,   {Raine} D.~J.,  2002, {Accretion Power in
  Astrophysics: Third Edition}

\bibitem[\protect\citeauthoryear{{Freitas} et~al.,}{{Freitas}
  et~al.}{2018}]{freitas18}
{Freitas} I.~C.,  et~al., 2018, \mndoi [\mnras] {10.1093/mnras/sty303}, \href
  {http://adsabs.harvard.edu/abs/2018MNRAS.476.2760F} {476, 2760}

\bibitem[\protect\citeauthoryear{{Hicks}, {Davies}, {Maciejewski}, {Emsellem},
  {Malkan}, {Dumas}, {M{\"u}ller-S{\'a}nchez}  \& {Rivers}}{{Hicks}
  et~al.}{2013}]{hicks13}
{Hicks} E.~K.~S.,  {Davies} R.~I.,  {Maciejewski} W.,  {Emsellem} E.,  {Malkan}
  M.~A.,  {Dumas} G.,  {M{\"u}ller-S{\'a}nchez} F.,   {Rivers} A.,  2013,
  \mndoi [\apj] {10.1088/0004-637X/768/2/107}, \href
  {http://adsabs.harvard.edu/abs/2013ApJ...768..107H} {768, 107}

\bibitem[\protect\citeauthoryear{{Hook}, {J{\o}rgensen}, {Allington-Smith},
  {Davies}, {Metcalfe}, {Murowinski}  \& {Crampton}}{{Hook}
  et~al.}{2004}]{hook04}
{Hook} I.~M.,  {J{\o}rgensen} I.,  {Allington-Smith} J.~R.,  {Davies} R.~L.,
  {Metcalfe} N.,  {Murowinski} R.~G.,   {Crampton} D.,  2004, \mndoi [\pasp]
  {10.1086/383624}, \href {http://adsabs.harvard.edu/abs/2004PASP..116..425H}
  {116, 425}

\bibitem[\protect\citeauthoryear{{Jarrett}, {Chester}, {Cutri}, {Schneider}  \&
  {Huchra}}{{Jarrett} et~al.}{2003}]{2mass}
{Jarrett} T.~H.,  {Chester} T.,  {Cutri} R.,  {Schneider} S.~E.,   {Huchra}
  J.~P.,  2003, \mndoi [\aj] {10.1086/345794}, \href
  {http://adsabs.harvard.edu/abs/2003AJ....125..525J} {125, 525}

\bibitem[\protect\citeauthoryear{{Koribalski} et~al.,}{{Koribalski}
  et~al.}{2004}]{systevel}
{Koribalski} B.~S.,  et~al., 2004, \mndoi [\aj] {10.1086/421744}, \href
  {http://adsabs.harvard.edu/abs/2004AJ....128...16K} {128, 16}

\bibitem[\protect\citeauthoryear{{Lena}}{{Lena}}{2014}]{lena14}
{Lena} D.,  2014, preprint, \href
  {http://adsabs.harvard.edu/abs/2014arXiv1409.8264L} {} (\mn@eprint {arXiv}
  {1409.8264})

\bibitem[\protect\citeauthoryear{{Lena}, {Robinson}, {Storchi-Bergmann},
  {Couto}, {Schnorr-M{\"u}ller}  \& {Riffel}}{{Lena} et~al.}{2016}]{lena16}
{Lena} D.,  {Robinson} A.,  {Storchi-Bergmann} T.,  {Couto} G.~S.,
  {Schnorr-M{\"u}ller} A.,   {Riffel} R.~A.,  2016, \mndoi [\mnras]
  {10.1093/mnras/stw896}, \href
  {http://adsabs.harvard.edu/abs/2016MNRAS.459.4485L} {459, 4485}

\bibitem[\protect\citeauthoryear{{Markwardt}}{{Markwardt}}{2009}]{markwardt09}
{Markwardt} C.~B.,  2009, in {Bohlender} D.~A.,  {Durand} D.,   {Dowler} P.,
  eds,  Astronomical Society of the Pacific Conference Series Vol. 411,
  Astronomical Data Analysis Software and Systems XVIII. p.~251 (\mn@eprint
  {arXiv} {0902.2850})

\bibitem[\protect\citeauthoryear{{Martini}, {Regan}, {Mulchaey}  \&
  {Pogge}}{{Martini} et~al.}{2003}]{martini03}
{Martini} P.,  {Regan} M.~W.,  {Mulchaey} J.~S.,   {Pogge} R.~W.,  2003, \mndoi
  [\apjs] {10.1086/367817}, \href
  {http://adsabs.harvard.edu/abs/2003ApJS..146..353M} {146, 353}

\bibitem[\protect\citeauthoryear{{Medling} et~al.,}{{Medling}
  et~al.}{2015}]{medling15}
{Medling} A.~M.,  et~al., 2015, \mndoi [\mnras] {10.1093/mnras/stv081}, \href
  {http://adsabs.harvard.edu/abs/2015MNRAS.448.2301M} {448, 2301}

\bibitem[\protect\citeauthoryear{{Menezes}, {da Silva}, {Ricci}, {Steiner},
  {May}  \& {Borges}}{{Menezes} et~al.}{2015}]{menezes15}
{Menezes} R.~B.,  {da Silva} P.,  {Ricci} T.~V.,  {Steiner} J.~E.,  {May} D.,
  {Borges} B.~W.,  2015, \mndoi [\mnras] {10.1093/mnras/stv629}, \href
  {http://adsabs.harvard.edu/abs/2015MNRAS.450..369M} {450, 369}

\bibitem[\protect\citeauthoryear{{Osterbrock} \& {Ferland}}{{Osterbrock} \&
  {Ferland}}{2006}]{osterbrock06}
{Osterbrock} D.~E.,  {Ferland} G.~J.,  2006, {Astrophysics of gaseous nebulae
  and active galactic nuclei}

\bibitem[\protect\citeauthoryear{{Phillips}, {Charles}  \&
  {Baldwin}}{{Phillips} et~al.}{1983}]{philips83}
{Phillips} M.~M.,  {Charles} P.~A.,   {Baldwin} J.~A.,  1983, \mndoi [\apj]
  {10.1086/160797}, \href {http://adsabs.harvard.edu/abs/1983ApJ...266..485P}
  {266, 485}

\bibitem[\protect\citeauthoryear{{Revalski}, {Crenshaw}, {Kraemer}, {Fischer},
  {Schmitt}  \& {Machuca}}{{Revalski} et~al.}{2018}]{revalski18}
{Revalski} M.,  {Crenshaw} D.~M.,  {Kraemer} S.~B.,  {Fischer} T.~C.,
  {Schmitt} H.~R.,   {Machuca} C.,  2018, \mndoi [\apj]
  {10.3847/1538-4357/aab107}, \href
  {http://adsabs.harvard.edu/abs/2018ApJ...856...46R} {856, 46}

\bibitem[\protect\citeauthoryear{{Riffel}}{{Riffel}}{2010}]{profit}
{Riffel} R.~A.,  2010, \mndoi [\apss] {10.1007/s10509-010-0317-y}, \href
  {http://adsabs.harvard.edu/abs/2010Ap%26SS.327..239R} {327, 239}

\bibitem[\protect\citeauthoryear{{Riffel}}{{Riffel}}{2013}]{baaa13}
{Riffel} R.~A.,  2013, Boletin de la Asociacion Argentina de Astronomia La
  Plata Argentina, \href {http://adsabs.harvard.edu/abs/2013BAAA...56...13R}
  {56, 13}

\bibitem[\protect\citeauthoryear{{Riffel}, {Storchi-Bergmann}, {Winge}  \&
  {Barbosa}}{{Riffel} et~al.}{2006}]{eso428}
{Riffel} R.~A.,  {Storchi-Bergmann} T.,  {Winge} C.,   {Barbosa} F.~K.~B.,
  2006, \mndoi [\mnras] {10.1111/j.1365-2966.2006.11050.x}, \href
  {http://adsabs.harvard.edu/abs/2006MNRAS.373....2R} {373, 2}

\bibitem[\protect\citeauthoryear{{Riffel}, {Storchi-Bergmann}, {Riffel}  \&
  {Pastoriza}}{{Riffel} et~al.}{2010}]{mrk1066:pop}
{Riffel} R.~A.,  {Storchi-Bergmann} T.,  {Riffel} R.,   {Pastoriza} M.~G.,
  2010, \mndoi [\apj] {10.1088/0004-637X/713/1/469}, \href
  {http://adsabs.harvard.edu/abs/2010ApJ...713..469R} {713, 469}

\bibitem[\protect\citeauthoryear{{Riffel}, {Riffel}, {Ferrari}  \&
  {Storchi-Bergmann}}{{Riffel} et~al.}{2011}]{mrk1157:pop}
{Riffel} R.,  {Riffel} R.~A.,  {Ferrari} F.,   {Storchi-Bergmann} T.,  2011,
  \mndoi [\mnras] {10.1111/j.1365-2966.2011.19061.x}, \href
  {http://adsabs.harvard.edu/abs/2011MNRAS.416..493R} {416, 493}

\bibitem[\protect\citeauthoryear{{Riffel}, {Storchi-Bergmann}  \&
  {Winge}}{{Riffel} et~al.}{2013}]{mrk79}
{Riffel} R.~A.,  {Storchi-Bergmann} T.,   {Winge} C.,  2013, \mndoi [\mnras]
  {10.1093/mnras/stt045}, \href
  {http://adsabs.harvard.edu/abs/2013MNRAS.430.2249R} {430, 2249}

\bibitem[\protect\citeauthoryear{{Riffel}, {Storchi-Bergmann}  \&
  {Riffel}}{{Riffel} et~al.}{2015}]{n5929}
{Riffel} R.~A.,  {Storchi-Bergmann} T.,   {Riffel} R.,  2015, \mndoi [\mnras]
  {10.1093/mnras/stv1129}, \href
  {http://adsabs.harvard.edu/abs/2015MNRAS.451.3587R} {451, 3587}

\bibitem[\protect\citeauthoryear{{Riffel}, {Storchi-Bergmann}, {Riffel},
  {Dahmer-Hahn}, {Diniz}, {Sch{\"o}nell}  \& {Dametto}}{{Riffel}
  et~al.}{2017}]{Riffel_LLP_stel}
{Riffel} R.~A.,  {Storchi-Bergmann} T.,  {Riffel} R.,  {Dahmer-Hahn} L.~G.,
  {Diniz} M.~R.,  {Sch{\"o}nell} A.~J.,   {Dametto} N.~Z.,  2017, \mndoi
  [\mnras] {10.1093/mnras/stx1308}, \href
  {http://adsabs.harvard.edu/abs/2017MNRAS.470..992R} {470, 992}

\bibitem[\protect\citeauthoryear{{Sandage}}{{Sandage}}{1978}]{sandage78}
{Sandage} A.,  1978, \mndoi [\aj] {10.1086/112271}, \href
  {http://adsabs.harvard.edu/abs/1978AJ.....83..904S} {83, 904}

\bibitem[\protect\citeauthoryear{{Schmitt}, {Storchi-Bergmann}  \&
  {Baldwin}}{{Schmitt} et~al.}{1994}]{schmitt94}
{Schmitt} H.~R.,  {Storchi-Bergmann} T.,   {Baldwin} J.~A.,  1994, \mndoi
  [\apj] {10.1086/173802}, \href
  {http://adsabs.harvard.edu/abs/1994ApJ...423..237S} {423, 237}

\bibitem[\protect\citeauthoryear{{Schmitt}, {Donley}, {Antonucci}, {Hutchings},
  {Kinney}  \& {Pringle}}{{Schmitt} et~al.}{2003}]{schmitt03}
{Schmitt} H.~R.,  {Donley} J.~L.,  {Antonucci} R.~R.~J.,  {Hutchings} J.~B.,
  {Kinney} A.~L.,   {Pringle} J.~E.,  2003, \mndoi [\apj] {10.1086/381224},
  \href {http://adsabs.harvard.edu/abs/2003ApJ...597..768S} {597, 768}

\bibitem[\protect\citeauthoryear{{Simpson}, {Wilson}, {Bower}, {Heckman},
  {Krolik}  \& {Miley}}{{Simpson} et~al.}{1997}]{simpson97}
{Simpson} C.,  {Wilson} A.~S.,  {Bower} G.,  {Heckman} T.~M.,  {Krolik} J.~H.,
   {Miley} G.~K.,  1997, \mndoi [\apj] {10.1086/303466}, \href
  {http://adsabs.harvard.edu/abs/1997ApJ...474..121S} {474, 121}

\bibitem[\protect\citeauthoryear{{Thomas} et~al.,}{{Thomas}
  et~al.}{2017}]{thomas17}
{Thomas} A.~D.,  et~al., 2017, \mndoi [\apjs] {10.3847/1538-4365/aa855a}, \href
  {http://adsabs.harvard.edu/abs/2017ApJS..232...11T} {232, 11}

\bibitem[\protect\citeauthoryear{{Tody}}{{Tody}}{1986}]{tody86}
{Tody} D.,  1986, in {Crawford} D.~L.,  ed.,  \procspie Vol. 627,
  Instrumentation in astronomy VI. p.~733, \mndoi{10.1117/12.968154}

\bibitem[\protect\citeauthoryear{{Tody}}{{Tody}}{1993}]{tody93}
{Tody} D.,  1993, in {Hanisch} R.~J.,  {Brissenden} R.~J.~V.,   {Barnes} J.,
  eds,  Astronomical Society of the Pacific Conference Series Vol. 52,
  Astronomical Data Analysis Software and Systems II. p.~173

\bibitem[\protect\citeauthoryear{{Urry} \& {Padovani}}{{Urry} \&
  {Padovani}}{1995}]{up95}
{Urry} C.~M.,  {Padovani} P.,  1995, \mndoi [\pasp] {10.1086/133630}, \href
  {http://adsabs.harvard.edu/abs/1995PASP..107..803U} {107, 803}

\bibitem[\protect\citeauthoryear{{Wilson} \& {Tsvetanov}}{{Wilson} \&
  {Tsvetanov}}{1994}]{wilson94}
{Wilson} A.~S.,  {Tsvetanov} Z.~I.,  1994, \mndoi [\aj] {10.1086/116935}, \href
  {http://adsabs.harvard.edu/abs/1994AJ....107.1227W} {107, 1227}

\bibitem[\protect\citeauthoryear{{Wylezalek} et~al.,}{{Wylezalek}
  et~al.}{2017}]{dominika17}
{Wylezalek} D.,  et~al., 2017, \mndoi [\mnras] {10.1093/mnras/stx246}, \href
  {http://adsabs.harvard.edu/abs/2017MNRAS.467.2612W} {467, 2612}

\bibitem[\protect\citeauthoryear{{de Vaucouleurs}, {de Vaucouleurs}  \&
  {Corwin}}{{de Vaucouleurs} et~al.}{1976}]{dv76}
{de Vaucouleurs} G.,  {de Vaucouleurs} A.,   {Corwin} J.~R.,  1976, in Second
  reference catalogue of bright galaxies, Vol. 1976, p. Austin: University of
  Texas Press..

\makeatother
\end{thebibliography}
\end{document}